\documentclass[amsmath,amssymb,twocolumn,prl,showpacs,nofootinbib,floatfix,superscriptaddress]{revtex4}
\usepackage{graphics}
\usepackage{dcolumn}
\usepackage{bm}
\usepackage{longtable}
\usepackage[dvips]{graphicx}
\usepackage{dcolumn}
\usepackage{footnote}
\usepackage{amssymb}
\usepackage{multirow}
\begin{document}

\title{Direct measurements of $^{22}$Na($p,\gamma$)$^{23}$Mg resonances and consequences for $^{22}$Na production in classical novae}
\author{A. L. Sallaska}
\affiliation{Department of Physics, University of Washington,
Seattle, WA 98195-1560, USA}
\email[Corresponding author: ]{sallaska@u.washington.edu}
\author{C. Wrede}
\affiliation{Department of Physics, University of Washington,
Seattle, WA 98195-1560, USA}
\author{A. Garc\'{\i}a}
\affiliation{Department of Physics, University of Washington,
Seattle, WA 98195-1560, USA}
\author{D. W. Storm}
\affiliation{Department of Physics, University of Washington,
Seattle, WA 98195-1560, USA}
\author{T. A. D. Brown}
\altaffiliation{Present address: Mary Bird Perkins Cancer Center, Baton Rouge, LA 70809, USA}
\affiliation{Department of Physics, University of Washington,
Seattle, WA 98195-1560, USA}
\author{C. Ruiz}
\affiliation{TRIUMF, Vancouver, BC V6T 2A3, Canada}
\author{K. A.  Snover}
\affiliation{Department of Physics, University of Washington,
Seattle, WA 98195-1560, USA}
\author{D. F.  Ottewell}
\affiliation{TRIUMF, Vancouver, BC V6T 2A3, Canada}
\author{L. Buchmann}
\affiliation{TRIUMF, Vancouver, BC V6T 2A3, Canada}
\author{C. Vockenhuber}
\altaffiliation{Present address: Laboratory of Ion Beam Physics,
ETH Zurich, 8093 Zurich, Switzerland}
\affiliation{TRIUMF, Vancouver, BC V6T 2A3, Canada}
\author{D. A. Hutcheon}
\affiliation{TRIUMF, Vancouver, BC V6T 2A3, Canada}
\author{J. A. Caggiano}
\altaffiliation{Present address: Lawrence Livermore National Laboratory, Livermore, CA 94550, USA}
\affiliation{TRIUMF, Vancouver, BC V6T 2A3, Canada}

\date{\today}

\begin{abstract}
The radionuclide $^{22}$Na is a potential astronomical observable that is expected to be produced in classical novae in quantities that depend on the thermonuclear rate of the $^{22}$Na($p,\gamma$)$^{23}$Mg reaction. We have measured the strengths of low-energy $^{22}$Na($p,\gamma$)$^{23}$Mg resonances directly and absolutely using a radioactive $^{22}$Na target.  We find the strengths of resonances at $E_{\textrm{p}}  = 213$, 288, 454, and 610 keV to be higher than previous measurements by factors of 2.4 to 3.2, and we exclude important contributions to the rate from proposed resonances at $E_{\textrm{p}}  = 198$, 209, and 232 keV. The $^{22}$Na abundances expected in the ejecta of classical novae are reduced by a factor of $\approx$ 2.
\end{abstract}


\maketitle

A classical nova is a thermonuclear outburst on the surface of a white-dwarf star that is accreting hydrogen-rich material from a binary companion.  Such novae are ideal sites for the modeling of explosive nucleosynthesis because most of the relevant thermonuclear reaction rates are based on experimental information~\cite{il02apj,josenew}. Astronomical observations of novae enriched in neon at ultraviolet, optical, and infrared wavelengths reveal that chemical elements as heavy as calcium can be synthesized and ejected~\cite{ge98pas}, consistent with a model where the accreted layer mixes with neon seed material in the underlying white dwarf to synthesize heavier elements via a network of proton-induced reactions and $\beta$ decays~\cite{tr86apj}.


Orbiting gamma-ray telescopes have the potential to detect lines that characterize the decays of particular isotopes produced in novae, such as $^{26}$Al and $^{22}$Na.  While novae may be one of many sources for the steady-state Galactic $^{26}$Al abundance~\cite{jo99apj,di06nat,ru06prl}, its long half life (720,000 yr) precludes its correlation with particular events.  More detailed information on the underlying physical processes can be provided by the detection of Galactic $^{22}$Na~\cite{jo99apj}, for which novae are expected to be the primary source.  The 2.6 yr half life of $^{22}$Na is short enough that it will be localized, yet long enough that it will survive beyond the opaque explosive phase (and hence be visible)~\cite{cl74apj}, assuming it is not destroyed during the explosion.

Recent modeling of $^{22}$Na production in novae has focused on oxygen-neon (ONe) white dwarfs~\cite{jo99apj,hi03npa,bi03prl,je04prl}, which are more massive than common carbon-oxygen (CO) white dwarfs and provide a source of neon~\cite{tr86apj}. Models of ONe nova predict $^{22}$Na yields that are slightly lower than the upper limits set by the non-observation of the characteristic (1275-keV) $^{22}$Na line in neon-type novae in the Galactic disk~\cite{iy95aas}. A key variable in the comparison of models to observations is the destruction of $^{22}$Na in novae, which depends crucially on the $^{22}$Na($p,\gamma$)$^{23}$Mg reaction rate~\cite{jo99apj,il02apj,hi03npa,je04prl}.

The thermonuclear rate of the $^{22}$Na($p,\gamma$)$^{23}$Mg reaction at peak nova temperatures ($0.1 < T < 0.4$ GK) is dominated by narrow, isolated resonances, and the currently accepted rate is based on two previous sets of direct measurements using radioactive $^{22}$Na targets~\cite{se90npa,st96npa}.  The first set~\cite{se90npa} yielded strengths for resonances with laboratory proton energies $E_{\textrm{p}} \geq 290$ keV.  The second set~\cite{st96npa} identified a new resonance at 213 keV and provided a measurement of its strength relative to strengths from the first set~\cite{se90npa}, with the conclusion that it dominates the rate. In 2004, measurements of the $^{12}$C($^{12}$C,$n\gamma$)$^{23}$Mg reaction~\cite{je04prl} provided an indirect basis for possible new resonances including one at $E_{\textrm{p}} = 198$ keV, which the authors speculated could supersede the contributions of all other $^{22}$Na($p,\gamma$)$^{23}$Mg resonances.

We have made direct, absolute measurements of key $^{22}$Na($p,\gamma$)$^{23}$Mg resonance strengths for $E_{\textrm{p}} < 610$ keV and searched for the proposed 198-keV resonance at the Center for Experimental Nuclear Physics and Astrophysics (CENPA) of the University of Washington using $^{22}$Na targets implanted at the radioactive ion beam facility TRIUMF-ISAC. Unlike Ref.~\cite{se90npa}, we bombarded all target atoms by scanning the beam uniformly over the target area~\cite{ju10prc} and integrating the excitation function for each resonance.  Our method is not very sensitive to target non-uniformity and stoichiometry or to the evolution of the target distribution due to prolonged bombardment.  More details will be given in Refs.~\cite{sa10prc,sallaskathesis}.

We produced $^{22}$Na by bombarding thick SiC targets with a 40-$\mu$A, 500-MeV proton beam from the TRIUMF cyclotron.  A surface ionization source and a high resolution mass separator were used to yield a 10-nA, 30-keV beam of $^{22}$Na$^+$.  Three 300-$\mu$Ci targets of $^{22}$Na were fabricated by rastering the ion beam over a 5-mm diameter collimator and into an OFHC copper substrate.  To prevent sputtering of $^{22}$Na by the intense proton beam during the ($p,\gamma$) measurements, two targets were coated with 20 nm of Cr using vacuum evaporation~\cite{br09nim}. The Cr-coated targets proved to be much more stable than the bare targets and were used to acquire the data for all resonances, except the proposed one at $E_{\textrm{p}} = 232$ keV.

Proton beams of $\approx 40~\mu$A were produced at CENPA by accelerating protons from an ion source at the terminal of a tandem Van de Graaff accelerator.  The extracted beam was analyzed using a $90^{\circ}$ magnet monitored with an NMR probe.  The beam line was equipped with a magnetic raster, a target chamber with a dual cold-shroud system, three electrically isolated collimators in series (including one on a sliding ladder), an electron suppressor, and a water-cooled target mount.  Pressure in the chamber was held at $\approx 1.5\times 10^{-7}$ torr.  At each resonance energy the beam was tuned, with the raster off, through 3- and 1-mm diameter collimators on the sliding ladder backed by electrically isolated beam stops until transmissions of $\gtrsim 95\%$ and $\gtrsim 50\%$, respectively, were achieved.  After tuning, an 8-mm collimator open to the target was inserted, and the raster was used to irradiate the implanted target area uniformly during data acquisition.  The beam current on the target was integrated to measure the total number of incident protons, $N_b$.

Two detection systems were positioned at $\pm 55^{\circ}$ to the beam axis.  Each system consisted of a high purity Ge (HPGe) crystal surrounded by Pb shielding and scintillators for cosmic-ray rejection.  Pb plates 26-mm thick were positioned between the target and the HPGe detectors to reduce the 511-keV $\gamma$-ray detection rate.  Data acquisition was triggered by signals from the HPGe detectors, and HPGe signals in coincidence with scintillator signals were rejected.  Dead time was monitored using a pulser.  The beam position on the target was inferred for each event by measuring and recording the vertical and horizontal magnetic fields produced by the raster.

Measurements were made on known $^{22}$Na($p,\gamma$)$^{23}$Mg resonances~\cite{se90npa,st96npa}, which we find at $E_{\textrm{p}} = 213, 288, 454$, and 610 keV, and on the proposed resonances at 198, 209~\cite{je04prl}, and 232 keV~\cite{pe00plb}.  For each resonance, the beam energy was stepped over a range of $\approx 25$ keV, taking into account energy losses in the Cr and Cu ($\approx 4$ keV, total), and the depth distribution of $^{22}$Na in the target (of order 10 keV-FWHM).  The raster amplitude was scaled by the square root of the beam energy to maintain a constant illuminated area on the target. For diagnostic purposes, measurements were also made on a thick, extended $^{27}$Al target, on a thick 5-mm diameter $^{27}$Al disk embedded in OFHC copper, and on an ion-implanted $^{23}$Na target similar to the $^{22}$Na targets.

If the beam is swept uniformly over the implanted area of the target, the resonance strength $\omega\gamma$ may be determined using the equation,

\begin{equation}
\int Y \, dE = 2 \pi^2 \lambdabar^2_{\textrm{lab}} \frac{m+M}{M} \,N_T \,\rho_b \, \omega\gamma,
\label{main}
\end{equation}

\noindent where $Y$ is the $\gamma$-ray yield at a beam energy $E$, $\int Y \, \mathrm{d}E$ is the integral over the excitation function, $\lambdabar_{\textrm{lab}}$ is the proton reduced de Broglie wavelength, $m$ is the proton mass, $M$ is the target mass, $N_T$ is the number of target atoms, $\rho_b = \frac{\mathrm{d}N_b}{\mathrm{d}A}/(Q/e)$ is the beam density normalized to the total number of incident beam particles, and $\frac{\mathrm{d}N_b}{\mathrm{d}A}$ is the areal density of the beam.  


The proton beam energy was calibrated using thick\--target excitation functions for well-known $^{27}$Al($p,\gamma$) resonances at $E_{\textrm{p}} = 326.6$, 405.5, 504.9, and 506.4 keV~\cite{en98npa}. The $\gamma$-ray energy spectra were calibrated using the 7357.84- and 5088.05-keV~\cite{me75npa} $\gamma$ rays from the 406-keV $^{27}$Al($p,\gamma$) resonance and their first-escape peaks.  Both of these calibrations were used to identify $^{22}$Na($p,\gamma$)$^{23}$Mg resonances, and they provided two independent measurements of the resonance and excitation energies.  The weighted averages of these values were adopted (Table~\ref{table}).

Detector efficiencies were determined for our setup, including the lead shielding, from absolute measurements at $E_{\gamma} =1332$ keV using a $^{60}$Co source calibrated to $1.7\%$
(99\% $C.L.$), and relative measurements using $^{24}$Na($\beta\gamma$) decay branches ($E_{\gamma} = 1369, 2754$ keV) and $^{27}$Al($p,\gamma$) reaction branches~\cite{en90npa,me75npa} at $E_{\textrm{p}} = 663$ keV ($E_{\gamma} =7575, 10451$ keV) and $E_{\textrm{p}} = 992$ keV ($E_{\gamma} =1779, 4742, 10762$ keV), together with detailed efficiency simulations using the Monte Carlo code PENELOPE~\cite{penelope}. The photopeak detection efficiency ranged from $4.7 \times 10^{-4}$ to $2.1 \times 10^{-4}$ in the relevant energy range of 4.2 to 8.2 MeV, with an uncertainty of $6\%$, which includes an estimated $3\%$ uncertainty arising from possible $^{22}$Na($p,\gamma$)$^{23}$Mg anisotropy.

\begin{figure}
\hfil\scalebox{.45}{\includegraphics{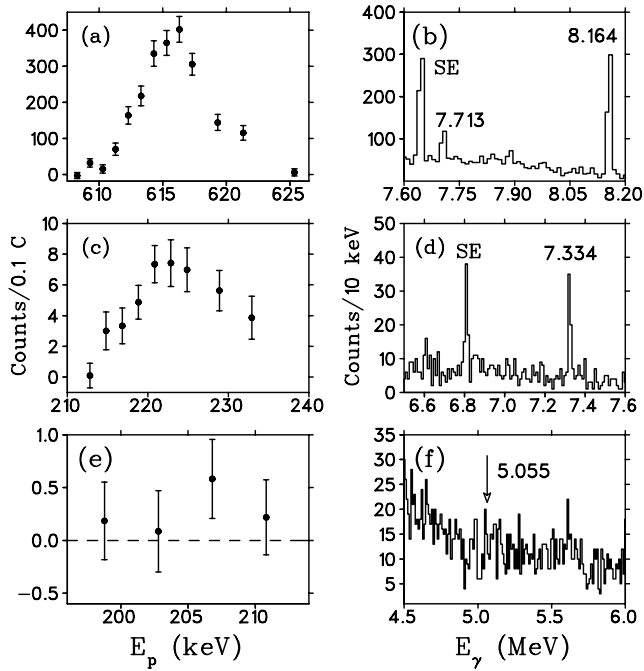}}\hfil
\caption{\label{yield} Excitation functions (left column) and corresponding summed $\gamma$-ray spectra (right column).   Panels (a) and (b) are for $E_{\textrm{p}} = 610$ keV, panels (c) and (d) are for $E_{\textrm{p}} = 213$ keV, and panels (e) and (f) are for $E_{\textrm{p}} = 198$ keV.   SE indicates a single-escape peak, and the arrow in (f) indicates where a potential $\gamma$ ray would be expected.}
\end{figure}

The yield for each $\gamma$-ray branch at each proton energy was determined by integrating the photopeak and (in selected cases) the first-escape peak. The integrated beam current, the detection efficiency, a 55 to $65\%$ correction for dead time, and a background subtraction were incorporated into this integration. The dead-time correction was verified by measuring $^{27}$Al($p,\gamma$) thick-target yields with and without a strong $^{22}$Na source in the vicinity of the detectors. $\int Y \, dE$ was determined from the excitation function for each potential primary branch using energy windows of width $\sim 25$ and $\sim 40$ keV for the two HPGe detectors, integrating over proton energy, and summing over the branches (Fig.~\ref{yield}). A $(10 \pm 5)\%$ correction was applied for the 213-keV resonance using the known shape of the excitation function because it was not completely covered by our data (Fig. \ref{yield} (c)). $\gamma$ rays with energies below $\approx 4$ MeV could not be measured due to intense backgrounds from the radioactive targets.

The initial value of $N_T$ was determined for each $^{22}$Na target by an \emph{in-situ} measurement of target activity using the 1275-keV line.  $^{22}$Na loss due to sputtering during proton bombardment~\cite{br09nim} was monitored by measuring strong resonances repeatedly and measuring residual $^{22}$Na activity in the beam line after the removal of each target.  The amount of potential loss for each measurement varied but was always $\leq 12\%$ for Cr-coated targets.

The quantity $\rho_b$ was measured by comparing the yields from the $^{27}$Al($p,\gamma$) reaction using the thick, extended $^{27}$Al target and the 5-mm disk $^{27}$Al target.  The resonances at 406 and 992 keV were used, yielding two values of $\rho_b$ in excellent agreement, whose average was adopted as the central value.  To estimate a systematic uncertainty for $\rho_b$, this measurement was supplemented by tests in which the raster amplitudes and collimator diameters were varied and by using a Monte-Carlo simulation that modeled the transport of the beam through the final components of the beam line. The result is $\rho_b = 2.58 \pm 0.25$ cm$^{-2}$, where the uncertainty incorporates potential non-uniformities in $\rho_b$ \emph{and} target density.

Using Eq.~\ref{main}, values of $\omega\gamma$ were extracted (Table~\ref{table}).  We find the strengths of the known $^{22}$Na($p,\gamma$)$^{23}$Mg resonances at $E_{\textrm{p}}  =$ 288, 454, and 610 keV to be higher than previous measurements~\cite{se90npa,st96npa} by a factor of 2.4 and that of the 213-keV resonance to be higher by a factor of 3.2.  Our branches~\cite{sa10prc} are roughly in agreement with Ref.~\cite{st96npa}.  We set upper limits on the strengths of the proposed resonances at 198, 209 keV~\cite{je04prl}, and 232 keV~\cite{pe00plb} by searching for known branches.  To extract upper limits, the shape of the excitation function from a reference resonance ($E_{\textrm{p}}  =$ 454 or 610 keV) was fit to the data of each proposed resonance, after stretching (set by the ratio of stopping powers), shifting, and normalizing the reference shape.  The shift and normalization parameters were allowed to vary, and the upper limit was extracted to be the 68\% quantile of the probability density function corresponding to the normalization.



Our direct limit for the 198-keV resonance improves upon the indirect limit from Ref.~\cite{je04prl} by a factor of 8.  We have set the first direct experimental limit  for the proposed~\cite{je04prl} resonance at 209 keV.  Our direct limit  for the 232-keV resonance is lower than the finite value of $2.2 \pm 1.0$ meV reported previously~\cite{pe00plb}, based on the $\beta$-delayed proton decay of $^{23}$Al.

We tested our experimental method by making an absolute
measurement of the 512-keV $^{23}$Na($p,\gamma$) resonance strength using the ion-implanted $^{23}$Na target. We measured the strength to be $79 \pm 17$ meV, consistent with the standard value of $91.3 \pm 12.5$ meV~\cite{il01apj}.

\begin{table}
\caption{Energies and strengths for $^{22}$Na($p,\gamma$)$^{23}$Mg resonances investigated in the present work ($Q$-value = 7580.53 $\pm$ 0.79 keV~\cite{23mgmass,muk}).  Present uncertainties and limits are at the $68\%~C.L.$  Finite strengths are the sum of partial strengths from measured branches, reported in detail elsewhere~\cite{sa10prc,sallaskathesis}.  }
\begin{ruledtabular}
\begin{tabular}{cccc}
Present &Present& Previous & Present\\
 $E_{\textrm{p}}$ (keV) &  $E_{\textrm{x}}$ (keV)& $\omega\gamma$ (meV) & $\omega\gamma$ (meV) \\
\colrule
 198\footnotemark[1]  & 7769\footnotemark[1]& $\leq 4.0$\footnotemark[2]      & $\leq 0.51$\footnotemark[5]  \\
 209\footnotemark[1]   &7780\footnotemark[1]& 0.05\footnotemark[2]            & $\leq 0.40$\footnotemark[6]  \\
 $213.4\pm1.4$         &7784.6 $\pm$ 1.2& $1.8\pm0.7$\footnotemark[3]     & $5.7^{+1.6}_{-0.9}$          \\
 232\footnotemark[1]  &7801\footnotemark[1] & $2.2\pm1.0$\footnotemark[4]     & $\leq 0.67$\footnotemark[7]  \\
 $288.0\pm1.1$        &7856.1 $\pm$ 1.0 & $15.8\pm3.4$\footnotemark[3]    & $39 \pm 8$                   \\
 $454.2\pm0.8$         &8015.3 $\pm$ 0.8& $68\pm20$\footnotemark[3]       & $166 \pm 22$                 \\
 $609.8\pm0.8$         &8163.9 $\pm$ 0.8& $235\pm33$\footnotemark[3]      & $591^{+103}_{-74}$           \\
\end{tabular}
\end{ruledtabular}
\footnotetext[1]{Nominal value from Refs.~\cite{je04prl,pe00plb} (not observed).}
\footnotetext[2]{Indirect estimate from Ref.~\cite{je04prl}.}
\footnotetext[3]{Direct measurement from Refs.~\cite{se90npa,st96npa}.}
\footnotetext[4]{Indirect estimate from Ref.~\cite{pe00plb}.}
\footnotetext[5]{Based on a search for the 5055-keV (58\%) branch~\cite{je04prl}.}
\footnotetext[6]{Based on a search for the 5067-keV (66\%) branch~\cite{je04prl}.}
\footnotetext[7]{Based on searches for the 7350-keV (66.4\%) and 7801-keV (30.0\%) branches~\cite{ia06prc}.}
\label{table}
\end{table}

\begin{figure}
\hfil\scalebox{.45}{\includegraphics{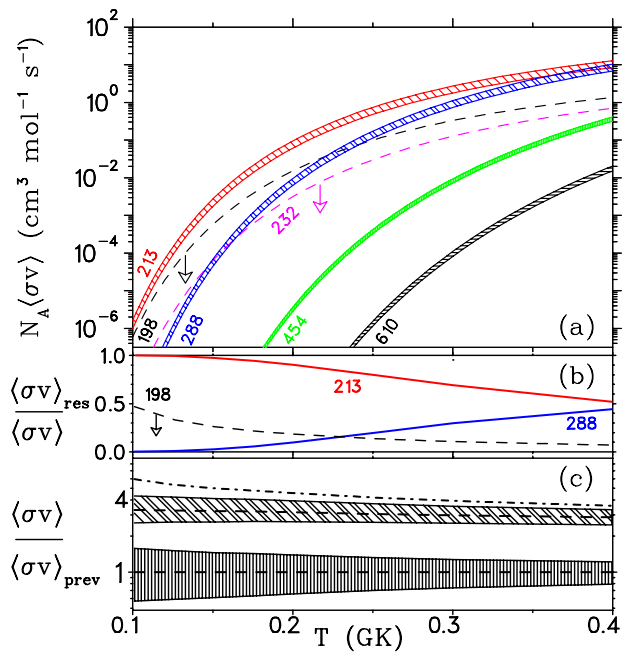}}\hfil
\caption{\label{fig:rate} Panel (a) shows contributions to the thermonuclear $^{22}$Na($p,\gamma$)$^{23}$Mg reaction rate from individual resonances labeled by $E_{\textrm{p}}$ in keV, based on the present measurements. Panel (b) shows the fractional contributions of selected resonances to the total present rate.  Upper limits (68\% $C.L.$) are denoted by dashed lines with arrows in (a) and (b). Panel (c) shows the ratios of reaction rates and their $1\sigma$ uncertainty limits to the central values of the reaction rate deduced from data in Refs.~\cite{se90npa,st96npa} on a log scale. The rates from Ref.~\cite{se90npa,st96npa} and the present work are represented by dashed lines in the vertical- and diagonal-hatched regions, respectively.  Including the limit for the 198-keV resonance increases the present upper limit to the dot-dashed line.}
\end{figure}

Contributions to the thermonuclear $^{22}$Na($p,\gamma$)$^{23}$Mg reaction rate were calculated using the presently measured resonance energies and strengths in Table~\ref{table} (Fig.~\ref{fig:rate} (a)).  Our upper limit for the 198-keV resonance and our strengths for the previously established resonances show that it is the 213-keV resonance that dominates the rate at peak novae temperatures ((Fig.~\ref{fig:rate} (b)).  The 288-keV resonance begins to make a substantial contribution near the highest nova temperature.  Due to the higher strengths for the 213- and 288-keV resonances, the total reaction rate is significantly higher than the currently accepted rate~\cite{se90npa,st96npa} by roughly a factor of 3 (Fig.~\ref{fig:rate} (c)).

Since the $^{22}$Na$(p,\gamma)^{23}$Mg reaction is not expected to have a significant effect on the total energy generation in novae, we use the results of one-zone post-processing network calculations~\cite{il02apj,hi03npa} to test the effects of our rate on $^{22}$Na production in nova models.  We estimate that the amount of $^{22}$Na produced is reduced by factors of 2 to 3 compared to the previous rate~\cite{se90npa,st96npa}, depending on the nova model used and the mass and composition of the underlying white dwarf. Treating the mixing of zones with the hydrodynamic code SHIVA~\cite{jose98} indicates that these factors will be diluted by $\approx 25\%$~\cite{privatecommunication,sa10prc}.  

In conclusion, we have measured low-energy $^{22}$Na($p,\gamma$)$^{23}$Mg resonance strengths using radioactive ion-implanted $^{22}$Na targets and a technique that is more reliable than those employed in previous measurements~\cite{se90npa,st96npa}.  We find the strengths of the key resonances that destroy $^{22}$Na in novae to be higher than the previous measurements~\cite{se90npa,st96npa} by factors of 2.4 to 3.2.  The contributions of proposed resonances at 198, 209, and 232 keV to the destruction of $^{22}$Na are found to be small, or negligible, at peak nova temperatures, and we have reduced uncertainties in the reaction rate.  In summary, our measurements show that $^{22}$Na will be destroyed much more efficiently in novae than previously thought, significantly reducing the prospects for the observation of $^{22}$Na via the characteristic 1275-keV $\gamma$-ray line using the spectrometer SPI onboard the currently-deployed INTEGRAL mission~\cite{he04nar}.

We gratefully acknowledge the contributions of J. Jos\'{e}, K. Deryckx, B.~M. Freeman, D. Short, the technical staffs at CENPA and TRIUMF, and the Athena cluster.  This work was supported by the U.S. Department of Energy under contract No. DE-FG02-97ER41020, the Natural Science and Engineering Research Council of Canada, and the National Research Council of Canada.

\end{document}